\documentclass[10pt,conference,twocolumn]{IEEEtran}
\IEEEoverridecommandlockouts

\usepackage{ifthen}

\usepackage[tbtags]{amsmath} 
\usepackage{amssymb}  
\usepackage{verbatim} 
\usepackage{amsxtra}  

\usepackage{mathabx}

\usepackage{enumerate}

\usepackage{amsfonts}
\usepackage{color}

\usepackage{subcaption}

\usepackage{wasysym}

\usepackage{cite}

\usepackage{letltxmacro}

\def\DHLhksqrt#1#2{%
\setbox0=\hbox{$#1\sqrt{#2\,}$}\dimen0=\ht0
\advance\dimen0-0.2\ht0
\setbox2=\hbox{\vrule height\ht0 depth -\dimen0}%
{\box0\lower0.4pt\box2}}

\usepackage{mathtools}
\mathtoolsset{showonlyrefs=true}

\newcommand{\isi}[1]{}
\newcommand{\isiincl}[2]{}

\newcommand{\googleincl}[2]{}
\newcommand{\googleinclabs}[3]{}

\usepackage{tikz}
\usetikzlibrary{shapes,arrows,decorations.markings,positioning}

\tikzstyle{plant} = [draw, fill=red!5, rectangle, 
    minimum height=3em, minimum width=6em]
\tikzstyle{block} = [draw, fill=blue!5, rectangle, 
    minimum height=3em, minimum width=6em]
\tikzstyle{sum} = [draw, fill=yellow!10, circle, node distance=1cm]
\tikzstyle{coord} = [coordinate]
\tikzstyle{gain} = [draw, fill=red!5, regular polygon, regular polygon sides=3, shape border rotate=-90]
\tikzstyle{pinstyle} = [pin edge={to-,thick,black}]

\tikzstyle{BitPipe} = [thick, decoration={markings,mark=at position
   1 with {\arrow[semithick]{open triangle 60}}},
   double distance=1.4pt, shorten >= 5.5pt,
   preaction = {decorate},
   postaction = {draw,line width=1.4pt, white,shorten >= 4.5pt}]

\usepackage{epsfig, graphicx, psfrag}

\usepackage[mathcal]{euscript}
\usepackage[cal=boondox]{mathalfa}

\usepackage{amsthm}

\newtheorem{thm}{Theorem}

\newtheorem{lem}{Lemma}

\newtheorem{corol}{Corollary}

\theoremstyle{definition}
\newtheorem{defn}{Definition}
\newtheorem*{defn*}{Definition}

\newtheorem*{scheme*}{Scheme}

\theoremstyle{remark}
\newtheorem{remark}{Remark}

\providecommand{\thmref}[1]{Th.~\ref{#1}}

\providecommand{\defnref}[1]{Def.~\ref{#1}}
\providecommand{\secref}[1]{Sec.~\ref{#1}}
\providecommand{\lemref}[1]{Lem.~\ref{#1}}

\providecommand{\remref}[1]{Rem.~\ref{#1}}
\providecommand{\figref}[1]{Fig.~\ref{#1}}
\providecommand{\colref}[1]{Corol.~\ref{#1}}

\newcommand{\ie}{i.e.}
\newcommand{\iid}{i.i.d.}

\newcommand{\cf}{cf.}

\newcommand{\reals}{\mathbb{R}}

\DeclareMathOperator{\plog}{log^+}

\let\limsup\relax
\DeclareMathOperator*{\limsup}{\overline{lim}}

\newcommand{\mG}{\mathcal{G}}
\newcommand{\mF}{\mathcal{F}}

\newcommand{\Comment}[1]{}
\newcommand{\old}[1]{}
\newcommand{\rem}[1]{}

\newcommand{\hx}{\hat{x}}

\newcommand{\tD}{\tilde D}

\newcommand{\Norm}[1]{\left\| #1 \right\|}

\providecommand{\comment}[1]{}

\providecommand{\norm}[1]{\Norm{#1}}

\newcommand{\beqn}[1]{\begin{eqnarray}\label{#1}}
\newcommand{\eeqn}{\end{eqnarray}}
\newcommand{\beq}[1]{\begin{equation}\label{#1}}
\newcommand{\eeq}{\end{equation}}

\providecommand{\half}{\frac{1}{2}}

\providecommand{\half}{\frac{1}{2}}

\makeatletter
\newcommand{\vast}{\bBigg@{4}}
\newcommand{\Vast}{\bBigg@{5}}
\makeatother

\newcommand{\opI}{I}
\providecommand{\causalKer}[2]{P(#1 \upupharpoons #2)}

\providecommand{\MI}[2]{{\opI \left( #1 ; #2 \right)}}
\providecommand{\DI}[2]{{\opI \left( #1 \to #2 \right)}}

\providecommand{\CMI}[3]{{\opI \left( #1 ; #2 \middle| #3 \right)}}
\providecommand{\CDI}[3]{{\opI \left( #1 \to #2 \upupharpoons #3 \right)}}
\providecommand{\CKL}[3]{{\bbD \left( #1 \middle\| #2 \middle| #3 \right)}}
\renewcommand{\H}[1]{{H \left( #1 \right)}}

\providecommand{\CH}[2]{{H \left( #1 \middle| #2 \right)}}

\providecommand{\bbD}{\mathbb{D}}
\providecommand{\bbE}{\mathbb{E}}

\providecommand{\E}[1]{\bbE \left[ #1 \right]}

\providecommand{\PACKET}{a}

\providecommand{\tind}{t}

\providecommand{\Rt}{R_\tind}

\providecommand{\markov}{\text{ --- }}

\usepackage{ifthen}
\usetikzlibrary{circuits.ee.IEC}
\usepackage{circuitikz}

\usepackage{mathtools}
\usepackage{MnSymbol}
\usepackage{comment}

\mathtoolsset{showonlyrefs=true}

\usepackage{comment}

\newcommand{\VersionLength}{short}

\providecommand{\ver}{\ifthenelse{\equal{\VersionLength}{long}}}
\providecommand{\nver}{\ifthenelse{\equal{\VersionLength}{short}}}

\providecommand{\figref}[1]{Fig.~\ref{#1}}
\providecommand{\secref}[1]{Sec.~\ref{#1}}

\ver{}{
	\setlength{\abovedisplayskip}{3.5pt}
	\setlength{\belowdisplayskip}{4pt}

}

\ver{\usepackage{subfig}}{}
\ver{\usepackage{hyperref}}{\usepackage{url}}

\begin{document}
    \title{Gauss--Markov Source Tracking 
    with \\ Side Information: Lower Bounds}
	
	\author{Omri Lev and Anatoly Khina
        \thanks{This work has received funding from the European Union's Horizon 2020 research and innovation programme under the Marie Sk\l odowska-Curie grant agreement No 708932.
        The work of O.\ Lev was supported in part by the Yitzhak and Chaya Weinstein Research Institute  for Signal Processing.}
    \\ {\tt \{omrilev@mail,anatolyk@eng\}.tau.ac.il}
    \\ School of Electrical Engineering, Tel Aviv University, Tel Aviv, Israel
	}
	\maketitle
	
	
	\begin{abstract}
%
		We consider the problem of causal source coding and causal decoding of a Gauss--Markov source, where the decoder has causal access to a side-information signal.
        We define the information causal rate--distortion function with causal decoder side information
		and prove that it bounds from below its operational counterpart. We further explain how to adapt the result to the setting of control over communication channels.
	\end{abstract}
	
	\ver{
	}{}
	
	\allowdisplaybreaks

\section{Introduction}
\label{s:intro}

Motivated by recent advances in tracking and control over networks \cite{SilvaDerpichOstergaard:ECDQ4Control,SahaiMitterPartI, DerpichOstergaard:ECDQ4CausalRDF,TatikondaSahaiMitter,StreamingWithFB:CNS,StavrouOstergaardCharalambousDerpich:ITW2017,BansalBasar:JSCC4Control,LQGoverAWGN:linear:NoBraslavsky,NairEvans:Quantization,MatveevSavkin:LQR-channel:SICON2007,FixedRateLQG:CDC2017,Yuksel:AC2014:LQG:separation,Yuksel:FixedRateControl,YukselBasarBook,KostinaHassibi:RDF4Control:AC,MineroFranceschettiDeyNair,JSCC4Control:AC2019}, we consider the setting where a decoder observes the system state corrupted by noise via an internal sensor, while it also receives quantized descriptions of the observations of the state from an external sensor over a rate-limited link.

We focus in this paper on the tracking (estimation) problem of a Gauss--Markov source over a rate-limited channel, \ie, causal encoding and decoding of the source; we view the internal noisy measurements of the state as side information that is available to the decoder
but not to the encoder. 

The idea of causal rate--distortion function (CRDF) was introduced in \cite{GorbunovPinsker:CausalRDF:Gauss}, where
\cite{TatikondaPhD,TatikondaSahaiMitter,Charalambous:DirectedInfo-GeneralCost:CDC2011} (see also \cite{KostinaHassibi:RDF4Control:AC,Tanaka:DirectedInfoLQG})
drew the connection between the CRDF and tracking of a Gauss--Markov source
over rate-limited links with causal encoding and decoding. 
Recently, two notable efforts have been made in determining bounds on the performance of these settings in the presence of decoder SI \cite{KostinaHassibi:SideInformationTracking,Starvou:GaussRDF_SI}, 
which provide a comprehensive set of definitions and bounds for this problem, 
by relying on the seminal work of Wyner and Ziv \cite{Wyner78,WynerZiv76} for rate--distortion with non-causal SI at the decoder.
However, since the technique of Wyner and Ziv relies on non-causal knowledge of the SI at the decoder, 
applying it for scenarios with causal SI imposes an additional slack when used to bound from below the operational CRDF with (\textit{causal}) SI, on top of the existing gap between the information and operational CRDFs without SI that stems from the causal encoding restriction \cite{LinderZamirCausal,NeuhoffGilbert:CausalSourceCodes:IT1982}.

Our goal in this paper is twofold: first, 
providing short proofs of the lower bounds in \cite{Starvou:GaussRDF_SI} via a simple observation; secondly, deriving a tighter lower bound on the performance of causal source coding with decoder SI that is strictly
higher than the bounds in \cite{KostinaHassibi:SideInformationTracking,Starvou:GaussRDF_SI}.
To derive the latter, we build on the work of Weissman and El Gamal \cite{WeissmanElGamal_FiniteLookAhead} for rate–-distortion with \textit{causal} SI and extend their results for CRDFs.


As a by product, we settle a conjecture in the negative by Stavrou and Skoglund \cite{Starvou:GaussRDF_SI} regarding the optimality of Wyner--Ziv-type CRDF bounds for causal tracking over additive white Gaussian noise (AWGN) channels, by proving that an adaptation of our new lower bound is strictly higher for~this~setting.
The rest of the paper is organized as follows. In \secref{s:Problem Statement}, we formulate the problem of tracking a Gauss--Markov source over a rate-limited link for several different SI scenarios. We review classical results and tools that are used throughout this work in \secref{s:Background}. We review the CRDF scenario without SI in \secref{s: No SI background}, and with two-sided SI \secref{s: Two-Sided SI background}.
We provide simple proofs for the existing results along with new tighter bounds on the CRDF with decoder SI in \secref{s: Causal RDF}, and adopt this result to the setting of control over communication channels in \secref{s: Discussion}.. 
We evaluate the expression of the new bound for a Gaussian and modulo test channels
in \secref{s: Numerical Evaluation - Gaussian Test Channel}. 

	\section{Problem Statement}
	\label{s:Problem Statement}
	
	\begin{figure}[t]
	    \centering
    	{\center
    	\includegraphics[width=\columnwidth]{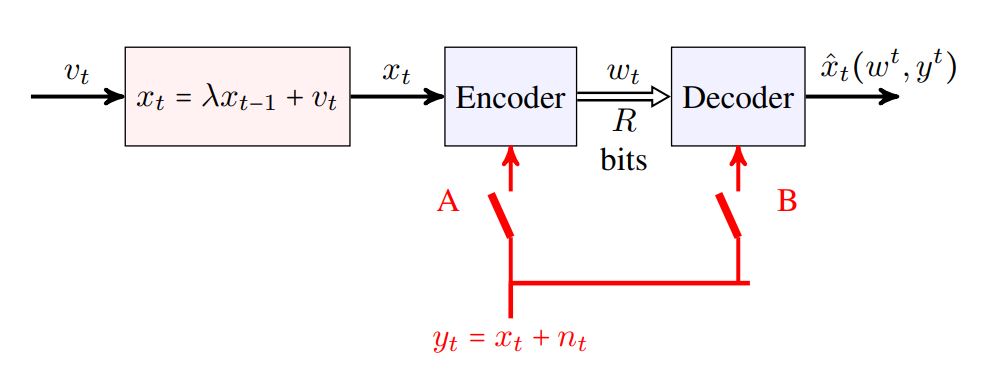}}
    	\caption{Scalar tracking system with driving WGN. The channel is a bit pipe with instantaneous rate constraint $R_{\tind}$. The presence of the SI in the encoder/decoder is according to the state of switch A/B, respectively. We assume that the SI is the original source $x_{\tind}$ after passing through a Gaussian channel.}
	    \label{fig: Tracking Model}
	\end{figure}   



  %
  %

  %
    
    In this section, we formalize the tracking setting treated in this work, depicted in \figref{fig: Tracking Model}.

\textit{Source.}
The source is 
generated by a first order Gauss--Markov model with zero initial condition ($x_0 = 0$):\footnote{The assumption $x_0 = 0$ can be easily replaced with a Gaussian $x_0$ that is independent of the system-disturbance sequence $\{ v_t \}$.}
\begin{align}
\label{eq:Source Gauss Markov}
    x_\tind &= \lambda x_{\tind-1} + v_{\tind} , & t = 1, \ldots, T ,
\end{align}
where $x_\tind \in \reals$ is the source sample at time $t$;
$v_{\tind}$ is the system disturbance at time $t$, 
whose temporal entries are independent and identically distributed (\iid) zero-mean Gaussian of variance $\sigma_v^2 > 0$; the eigenvalue $\lambda \in \reals$ is fixed and known.

\textit{Encoder.}
Observes the state $x_t$ at time $t$ and generates a packet 
$\PACKET_\tind \in \left\{1, \ldots, 2^{\Rt} \right\}$ of rate $\Rt$.


\textit{Channel.}
At time $\tind$, a packet 
$\PACKET_{\tind} \in \left\{ 1, 2, \ldots, 2^{R_{\tind}} \right\}$ 
is sent over a noiseless channel with rate $R_{\tind}$. 
The packets are subject to an average-rate constraint:\footnote{This is a more lenient constraint than the fixed-rate constraint. Consequently, our lower bounds are valid for both scenarios, although they might be too optimistic for the latter.} 
\begin{align}
    \label{eq:Overall Rate}
    	\frac{1}{T}\sum_{\tind=1}^{T} R_{\tind} \leq R.
\end{align}

\textit{Side information.}
The SI is a noisy version of the current source sample $x_{\tind}$, and is given by  
\begin{align}
\label{eq:Side Information}
    y_{\tind} = x_{\tind} + n_{\tind},
\end{align}
where $n_t$ is zero-mean Gaussian of variance $\sigma^{2}_{n}$, 
independent of $x^t$,\footnote{We denote temporal sequences by $a^t \triangleq \left( a_1, \ldots, a_t \right)$.}
and its temporal entries are \iid

\textit{Decoder.}
At time $t$, receives the packet $\PACKET_t$ and constructs an estimate $\hx_t$ of $x_{\tind}$.

    \textit{Distortion.}	
    The average quadratic distortion 
    at time $\tind$ is defined as
	\begin{align}
	\label{eq:Distortion}
	    D_{\tind} = \E{\left(x_{\tind} - \hat{x}_{\tind}\right)^{2}},
	\end{align}	
	and the average-stage distortion is defined as 
	\begin{align}
	\label{eq:Overall Distortion}
	    D = \frac{1}{T}\sum_{\tind = 1}^{T}D_{\tind}.
	\end{align}

\begin{defn}[Operational causal rate--distortion function]
\label{eq:CRDF:operational}
    The operational causal rate--distortion function (CRDF) $R_{c, op}(D)$
    is defined as the infimum of all achievable average rates $R$,  $\frac{1}{T}\sum_{\tind=1}^{T}R_{\tind} = R$, subject to an average distortion constraint $\frac{1}{T}\sum_{\tind = 1}^{T}D_{\tind}\leq D$.
\end{defn}

Different scenarios for the availability of the SI may be considered, corresponding to different states of switches A and B in \figref{fig: Tracking Model}:
\begin{itemize}
\item \textit{No SI (A open, B open).} 
    The encoder applies a \textit{causal function} $\mF_\tind$ to the source history $x^\tind$, to generate the packet $\PACKET_\tind \in \left\{1, \ldots, 2^{\Rt} \right\}$:
    $\PACKET_\tind = \mF_\tind \left( x^\tind \right)$,
    whereas the decoder 
    applies a \textit{causal function} $\mG_t$ to the sequence of received packets $\PACKET^t$, to construct an estimate $\hx_t$ of $x_{\tind}$:
    $\hx_{\tind} = \mG_{\tind}\left(\PACKET^t\right)$.

\item \textit{Two-sided SI (A closed, B closed).} 
    Here, both the encoder and the decoder have access to the SI and hence $\PACKET_\tind = \mF_\tind \left( x^t, y^t \right)$ and $\hx_{\tind} = \mG_{\tind}\left(\PACKET^t, y^t \right)$.
    
\item \textit{Decoder SI (A open, B closed).}
        Here, only the decoder has access to the SI. Thus, 
        $\PACKET_\tind = \mF_\tind \left( x^t \right)$ and $\hx_{\tind} = \mG_{\tind}\left(\PACKET^t, y^t \right)$.
\end{itemize}


\section{Background}
\label{s:Background}

\subsection{Batch Rate--Distortion}


In this section we review classical results from information theory on lossy compression.
The standard mode of operation assumes batch operation over long blocks ($T \to \infty$): The encoder observes a long block of source samples $x^T$, and maps them together to a (single) packet $\PACKET$; the decoder recovers the estimates $\hx^T$ of the the entire sequence upon receiving $\PACKET$, i.e., in a non-causal fashion [\cf~\eqref{eq:Distortion}].

Within this framework, information theory discriminates between four different scenarios of the availability of SI and its nature, which we present next for the commonly-considered case of an \textit{\iid}\ Gaussian source, corresponding to taking $\lambda = 0$ in \eqref{eq:Source Gauss Markov}:
\begin{itemize}
\item \textit{No SI.} 
    This is the classical rate--distortion scenario \cite{Shannon59:RDF}, \cite[Ch.~10]{CoverBook2Edition}. For which the rate--distortion function (RDF) is equal to 
    \begin{align}
    \label{eq:IT:noSI}
        R(D) = 
            \half \plog \frac{\sigma_v^2}{D}, 
    \end{align}
    where $\plog(x) \triangleq \max\{\log x, 0\}$.
    
\item \textit{Two-sided SI.} 
    This scenario can be recast as that of no SI with additional conditioning, 
    as both the encoder and the decoder know the SI. Thus, conditional RDF amounts to 
    \begin{align}
    \label{eq:IT:two-sided}
        R^\mathrm{both}(D) &= 
            \half \plog \frac{\sigma_{v|y}^2}{D} 
        =
            \half \plog \frac{\sigma_v^2\|\sigma_n^2}{D}, 
    \end{align}
    where $\sigma_{a|b}^2$ denotes the conditional variance of $a$ given $b$, and $a\|b \triangleq ab/(a+b)$.
    
\item \textit{Decoder non-causal SI.}
    Here, for the reconstruction of $x_t$ ($t\in \{1, \ldots, T\}$), the decoder may use the entire side information sequence $y^T$ in addition to $\PACKET$, whereas the encoder is oblivious of $y^T$.
    Surprisingly, a classical result due to Wyner \cite{Wyner78} (an adaptation to the Gaussian case of a result by Wyner and Ziv \cite{WynerZiv76}) states that, for an \iid\ Gaussian source, the RDF for this scenario, $R^\mathrm{NC}$, coincides with that of \eqref{eq:IT:two-sided}, i.e., 
    $R^\mathrm{NC}(D) \equiv R^\mathrm{both}(D)$.
    
\item \textit{Decoder causal SI.}
    This scenario is identical to the previous one except that now, 
    for the reconstruction $\hx_t$ of $x_t$ at time $t$, in addition to $\PACKET$,
    the decoder may use only the \textit{causal history} of the SI $y^t$.
    Weissman and El Gamal \cite{WeissmanElGamal_FiniteLookAhead} have shown that the RDF for this scenario is given~by\footnote{$a \markov b \markov c$ denotes a Markov chain, \ie, given $b$, $a$ is independent of $c$.}
    \begin{align}
    \label{eq:RDF:WeissmanElGamal}
        R^\mathrm{C}(D) &= 
        \inf_{\substack{ P(w|x) \::\: y \markov x \markov w ,
        \\  \E{\left(x - \hx(w,y) \right)^2} \leq D
        }} \MI{x}{w}
    \end{align}    

    and is higher than \eqref{eq:IT:two-sided}.
    Furthermore, it is bounded from above by
    \begin{align}
    \label{eq:RDF:Weissman-ElGamal}
        R^\mathrm{C}(D) &\leq \mathrm{c.e.} \left\{\half \plog \left(\frac{\sigma_v^2}{D} - \frac{\sigma_v^2}{\sigma_n^2}\right)\right\} 
     \triangleq \mathrm{c.e.} \left\{r(D) \right\}.
    \end{align}
    where c.e.\ denotes the \textit{convex envelope} operation, and is manifested by a straight line between the points $\left( D_c, r(D_c )\right)$ and $(\sigma_n^2\|\sigma_v^2,0)$ in the regime $D \in (D_c, D_{\max})$, 
    where $D_c$ is the solution to the equation $r(D_{c}) = \left( D_{c} - \sigma_v^2 \| \sigma_n^2 \right) \frac{d}{dD} r(D)\big|_{D = D_c}$; 
    the convex envelope comes into play only when $D_c < \sigma_n^2 \| \sigma_v^2$, i.e., only when $\sigma_n^2 < \sigma_v^2$.
\end{itemize}

\begin{remark}
\label{rem:classical-RDF:losses}
    The RDFs for the different scenarios serve as an outer bound for finite $T$ and are attainable only in the limit of $T \to \infty$.
    However, as have been proved by Zamir and Linder \cite{LinderZamirCausal}, 
    even in the limit of $T \to \infty$ (and even for \iid\ Gaussian sources) they are not attainable, in general (although they can be approached up to a fixed additive loss \cite[Ch.~5]{ZamirBook}).
    Finally, note that for the batch setting these results may be extended beyond the \iid\ setting ($\lambda \neq 0$); see \cite{ZamirKochmanErez:DPCM:IT,AnalogMatching}.
\end{remark}

\begin{remark}
\label{rem:classical-RDF:two-sided SI}
    When the side information is known to both the encoder and the decoder, 
    it turns out that the RDFs coincide for the cases when the SI is known causally and non-causally. Therefore, we do not distinguish between these two scenarios.
\end{remark}


\subsection{Directed Information}

The \textit{Directed Information} (DI) notion, introduced by Massey \cite{Massey:DirectedInfo}, 
is the causal counterpart of the classical \textit{Mutual Information} MI and is defined as follows. 
\begin{defn}[DI]
    The DI between $x^T$ and $y^T$ is defined as 
    \begin{subequations}
    \label{DI:def}
    \noeqref{DI:def:MI,DI:def:KL}
    \begin{align}
        \DI{x^T}{y^T} &= \sum_{t=1}^T \CMI{x^t}{y_t}{y^{t-1}} 
    \label{DI:def:MI}
     \\ &= \CKL{ \causalKer{y^T}{x^T} }{ P_{y^T} }{ P_{x^T} } ,
    \label{DI:def:KL}
    \end{align}
    \end{subequations}
    where $\CMI{\cdot}{\cdot}{\cdot}$ denotes the conditional MI, $\CKL{\cdot}{\cdot}{\cdot}$ is the conditional \textit{Kullback--Leibler divergence}, and 
    \begin{align}
    \label{eq:condKer}
        \causalKer{y^T}{x^T} \triangleq \prod_{\tind=1}^{T} P\left(y_{\tind}|y^{\tind-1},x^{\tind}\right)
    \end{align}
    is the \textit{causally conditional probability kernel} \ver{\cite[Ch.~3]{KramerPhD} (see
    also \cite{KostinaHassibi:SideInformationTracking})}{\cite[Ch.~3]{KramerPhD}, \cite{KostinaHassibi:SideInformationTracking}}.
\end{defn}

Clearly, $0 \leq \DI{x^T}{y^T} \leq \MI{x^T}{y^T}$, and for a sequence of independent pairs $\{(x_t,y_t)\}_{t=1}^T$, the DI and the MI coincide (see \cite[Ch.~3]{KramerPhD} for further details). 

The causally conditional DI is defined next and allows, in turn, to derive a chain-rule and a \textit{Data-Processing Inequality} (DPI) for DIs.
\begin{defn}
    The \textit{causally conditional DI} is defined as 
    \begin{align}
        \label{condDI:def}
        \CDI{x^T}{y^T}{z^T} &\triangleq \sum_{t=1}^T \CMI{x^t}{y_t}{y^{t-1},z^t}
        ,
    \end{align}
    and its lagged-by-one variant---as 
    \begin{align}
    \label{condDI_d:def}
        \CDI{x^T}{y^T}{z^{T-1}} 
        &\triangleq \sum_{t=1}^T \CMI{x^t}{y_t}{y^{t-1},z^{t-1}}. 
    \end{align}
\end{defn}

\begin{thm}[Chain rule for DIs \cite{Massey:DirectedInfo}, {\cite[Ch.~3]{KramerPhD}}]
\label{thm:chain rule-DI}
    \begin{subequations}
    \noeqref{eq:DI Chain Rule2}
    \begin{align}
        \label{eq:DI Chain Rule} 
        \DI{\left(x^T,y^T\right)}{z^T} &= \DI{x^T}{z^T} + \CDI{y^T}{z^T}{x^T}, \ 
     \\ \DI{x^T\!}{\left( y^T, z^T \right)} &= \CDI{x^T}{y^T\!}{\!z^{T-1}} + \CDI{x^T\!}{z^T\!}{y^T\!}. \ 
        \label{eq:DI Chain Rule2} 
    \end{align}
    \end{subequations}
\end{thm}

\begin{thm}[DPI for DIs \cite{SilvaDerpichOstergaard:ECDQ4Control,Tanaka:DirectedInfoLQG}]
\label{thm:DPI-DI}
Let $u^T, a^T, x^T$ satisfy the Markov relations $\left(x_t,a^{t-1}\right) \to \left(a^t,u^{t-1}\right) \to u_t$ for all $t \in \{1, 2, \ldots, T\}$.
Then,
\begin{align}
    \DI{x^{T}}{u^{T}} 
    \leq \CDI{x^T}{a^T}{u^{T-1}} .
\label{eq: DPI for DI} 
\end{align}
\end{thm}

\section{No SI}
\label{s: No SI background}

In this section we review known results for the scenario where SI is available to neither the encoder nor the decoder, corresponding to switches A and B being open in \figref{fig: Tracking Model}. 

\begin{defn}[\!\!\cite{GorbunovPinsker:CausalRDF:Gauss}]
    The information CRDF of a Gaussian source $\{x_t\}$ (without SI) is defined as 
    \begin{subequations}
    \label{eq: Classical Causal RDF}
    \noeqref{eq: Classical Causal RDF:Gauss}
    \begin{align}
        R_c(D) &= \limsup_{T \to \infty}
         \inf_{\substack{\causalKer{\hx^T}{x^T},
         \\ \frac{1}{T} \sum_{t}\E{ \norm{x_t - \hx_t}^2 } \leq D}}
         \frac{1}{T} \DI{x^T}{\hx^T} \qquad
    \label{eq: Classical Causal RDF:def}
     \\ &= \half \plog \frac{\lambda^2 D + \sigma_v^2}{D} \:.
    \label{eq: Classical Causal RDF:Gauss}
    \end{align}
    \end{subequations}
\end{defn}
\eqref{eq: Classical Causal RDF:Gauss} is derived in \cite{GorbunovPinsker:CausalRDF:Gauss,DerpichOstergaard:ECDQ4CausalRDF,Tanaka:UniformRateAllocation,StreamingWithFB:CNS}.

\begin{thm}
\label{thm:noSI:operational-information}
    The operational CRDF (without SI), $R_{c,op}(D)$, is bounded from below by the information CRDF (without SI)~\eqref{eq: Classical Causal RDF}:
    $R_c(D) \leq R_{c,op}(D)$.
\end{thm}
For a detailed proof see \cite{DerpichOstergaard:ECDQ4CausalRDF,StreamingWithFB:CNS}. 
\begin{remark}
\label{rem:noSI:loss}
    As mentioned in \remref{rem:classical-RDF:losses}, 
    equality in the lower bound of \thmref{thm:noSI:operational-information}
    \textit{cannot} be achieved, in general. Nonetheless, it can be mimicked up to a finite loss via entropy-coded dithered quantization \cite{SilvaDerpichOstergaard:ECDQ4Control,Tanaka:DirectedInfoLQG,StreamingWithFB:CNS}. Note, however, that this bound may become loose in the low-rate regime.
\end{remark}

\section{Two-Sided SI}
\label{s: Two-Sided SI background}

We now treat the two-sided SI scenario, \ie, the scenario in which the SI is available to both the encoder and the decoder, corresponding to both switches A and B being closed in \figref{fig: Tracking Model}.

\begin{defn}[Information CRDF with two-sided SI \cite{KostinaHassibi:SideInformationTracking}]
    The information CRDF with two-sided SI of a Gaussian source $\{x_t\}$
    with a jointly Gaussian SI $\{y_t\}$ that is known to both the encoder and the decoder is defined as
    \begin{subequations}
    \label{eq: Classical Causal RDF with SI}
    \noeqref{eq: Classical Causal RDF with SI:def,eq: Classical Causal RDF with SI:Gauss}
    \begin{align}
        R^\mathrm{both}_c(D) &= \limsup_{T \to \infty} 
         \inf_{\substack{ \causalKer{\hx^T}{x^T, y^T}, 
         \\ \frac{1}{T} \sum_t\E{ \norm{x_t - \hx_t}^2 } \leq D }}
         \frac{1}{T}\CDI{x^T}{\hx^T}{y^T} \quad\:
    \label{eq: Classical Causal RDF with SI:def}
     \\ &= \half \plog \frac{ \sigma^2_n \| (\lambda^2 D + \sigma^2_v) }{D} .
    \label{eq: Classical Causal RDF with SI:Gauss}
    \end{align}
    \end{subequations}
\end{defn}


\begin{thm}
\label{thm:two-sided:operational-information}
    The operational CRDF with two-sided SI, $R^\mathrm{both}_{c,op}(D)$, is bounded from below by 
    the information CRDF with two-sided SI \eqref{eq: Classical Causal RDF with SI}:
    $R^\mathrm{both}_c(D) \leq R^\mathrm{both}_{c,op}(D)$.
\end{thm}

    The setting with two-sided SI is equivalent to the no SI setting, w.r.t.\ to a (Gaussian) source that 
    is equal to $x_t$ given $y^t$.
    This simple observation allows a simple adaptation of the proof without SI to that of \thmref{thm:two-sided:operational-information}.

\begin{IEEEproof}
    By looking at the equivalent source $x_t|y^t$, the problem is equivalent to the no SI setting \eqref{eq: Classical Causal RDF:def}, 
    with the variance of the prediction error of $x_t$ given $\hx^{t-1}, y^t$ being 
    \begin{align}
    \label{eq: Prediction error easy}
        \sigma^{2}_{x_{\tind}|y^{\tind},\hx^{\tind-1}} 
         &= \left. \sigma^{2}_{y_{\tind}|x_{\tind}} \middle\| \sigma^{2}_{x_{\tind}|y^{\tind-1},\hx^{\tind-1}} \right. 
        = \sigma^2_n \| \left( \lambda^2 D_{\tind-1} + \sigma^2_v \right).
    \end{align} 	   
    
    Plugging it in \cite[Eq.~(18)]{StreamingWithFB:CNS} gives rise to 
    \begin{align}
    \label{eq: Two-Sided solution, 3}
        R^{both}_{c}(D) = \frac{1}{T}\sum_{t=1}^{T} \half \log \left(\sigma^2_n \| \left( \lambda^2 D_{t-1} + \sigma^2_v \right)\right)  - \half \log D_{t} \,.
    \end{align}	    
    
    By applying Jensen's inequality and taking $T \to \infty$ (i.e., repeating steps (18d),(18e) of \cite{StreamingWithFB:CNS}) we arrive at the desired result: 
    \begin{align}
    \label{eq: Two-Sided solution, 4}
        &R^{both}_{c}(D) = \half \log \frac{ \sigma^2_n \| \left( \lambda^2 D + \sigma^2_v \right) }{D} , 
    \end{align}	
    with $D$ being the average-stage distortion \eqref{eq:Overall Distortion}.
    Using \thmref{thm:noSI:operational-information} we conclude that $R^{both}_{c}(D) \leq R^\mathrm{both}_{c,op}(D)$.
\end{IEEEproof}


\section{Causal Rate--Distortion With Decoder SI}
\label{s: Causal RDF}

In this section, we treat the more involved scenario where the SI is known only to the decoder while the encoder is oblivious of the SI, corresponding to switch A being open and B begin closed in \figref{fig: Tracking Model}. 

We start by presenting a na\"ive lower bound.
\begin{lem}
\label{lem:decSI:twoSided - information}
    The operational CRDF with \textit{decoder SI}, $R^{dec}_{c,op}(D)$, is bounded from below by the information CRDF with \textit{two-sided SI} \eqref{eq: Classical Causal RDF with SI:def}:
    $R^{both}_{c}(D) \leq R^{dec}_{c,op}(D)$.
\end{lem} 
\begin{IEEEproof}
    Making the SI available (as a ``genie'') may only improve performance, and thus $R^\mathrm{both}_{c,op}(D) \leq R^{dec}_{c,op}(D)$. Using \thmref{thm:two-sided:operational-information}, the result follows.
\end{IEEEproof}

\begin{remark}
    Beyond the loss mentioned in Rems.~\ref{rem:classical-RDF:losses} and \ref{rem:noSI:loss} due to the causal encoding, 
    the lower bound in \lemref{lem:decSI:twoSided - information} is known to be loose even for the batch memoryless RDF setting \cite{WeissmanElGamal_FiniteLookAhead} due to the causal access to the SI at the decoder (see also \cite{Tuncel_ZeroDelayJSCCwithWZ}). 
\end{remark}

\begin{defn}[Information CRDF with decoder SI]
\label{def:Classical Causal RDF with SI @Rx}
    The information CRDF with decoder SI of a Gaussian source $\{x_t\}$
    with a jointly Gaussian SI $\{y_t\}$ that is known to the decoder is defined as
    \begin{align}
    \label{eq:CRDF:SI@DEC}
        R^\mathrm{dec}_c(D) &= \limsup_{T \to \infty} 
         \inf_{\substack{ \causalKer{w^T}{x^T}, \{\hx_t(w^t,y^t)\} : 
         \\ \frac{1}{T} \sum_{t}\E{ \norm{x_t - \hx_t}^2 } \leq D,
         \\ \left(y^t,x^{t-1}\right)\markov\left(x^t,w^{t-1}\right)\markov w_t}}
         \frac{1}{T} \DI{x^T}{w^T} . \quad\:
    \end{align}
\end{defn}

\begin{thm}
\label{thm:decSI:operational-information}
    The operational CRDF with decoder SI, $R_{c,op}^{dec}(D)$, is bounded from below by the information CRDF with decoder SI \eqref{eq:CRDF:SI@DEC}:
    $R^\mathrm{dec}_c(D) \leq R_{c,op}^{dec}(D)$.
\end{thm}

\begin{IEEEproof}
    We assume that the average distortion is equal to (or lower than) $D$ and bound the average rate $R$ (recall \defnref{eq:CRDF:operational}):
    %
    \begin{subequations}
    \label{eq:proof:CRDF:SI@DEC:converse}
    \begin{align}
        T R &\geq \H{\PACKET^T} 
    \label{eq:proof:CRDF:SI@DEC:converse:R>=H}
     \\ &= \sum_{t=1}^T \CH{\PACKET_t}{\PACKET^{t-1}}
    \label{eq:proof:CRDF:SI@DEC:converse:H-chain-rule}
     \\ &\geq \sum_{\tind=1}^T \CH{\PACKET_t}{\PACKET^{t-1}, w^{t-1}}
    \label{eq:proof:CRDF:SI@DEC:converse:cond reduces H}
     \\ &\geq \sum_{\tind=1}^T \CH{\PACKET_t}{a^{t-1}, w^{t-1}} 
        - \CH{\PACKET_t}{\PACKET^{t-1}, w^{t-1}, x^t} \quad
    \label{eq:proof:CRDF:SI@DEC:converse:H>=0}
     \\ &= \sum_{t=1}^T \CMI{x^t}{a_t}{a^{t-1}, w^{t-1}} 
    \label{eq:proof:CRDF:SI@DEC:converse:MI-def}
     \\ &= \CDI{x^T}{a^T}{w^{T-1}} 
    \label{eq:proof:CRDF:SI@DEC:converse:CDI-def}
     \\ &\geq \DI{x^T}{w^T} ,  
    \label{eq:proof:CRDF:SI@DEC:converse:DI-DPI}
     \\ &\geq T R^\mathrm{dec}_c(D) ,  
    \label{eq:proof:CRDF:SI@DEC:converse:CRDF}
    \end{align}
    \end{subequations}
    where 
    \eqref{eq:proof:CRDF:SI@DEC:converse:R>=H} follows from the problem statement, 
    \eqref{eq:proof:CRDF:SI@DEC:converse:H-chain-rule} is due to the chain rule for entropies, 
    \eqref{eq:proof:CRDF:SI@DEC:converse:cond reduces H} holds since conditioning does not increase entropy,
    \eqref{eq:proof:CRDF:SI@DEC:converse:H>=0} follows from the non-negativity of entropy,
    \eqref{eq:proof:CRDF:SI@DEC:converse:MI-def} and \eqref{eq:proof:CRDF:SI@DEC:converse:CDI-def} are by the definition of the conditional MI and lagged-by-one DI~\eqref{condDI_d:def}, respectively, 
    \eqref{eq:proof:CRDF:SI@DEC:converse:DI-DPI} follows from the DPI for DIs of \thmref{thm:DPI-DI} for $x^t,a^t,w^t$ satisfying the Markov relations
    \begin{align}
        \label{eq: Markov for proof}
        \left(x_t,a^{t-1}\right) \markov \left(a^t,w^{t-1}\right) \markov w_t 
    \end{align}
    for all $t \in \{1, 2, \ldots, T\}$ ($a_t \triangleq 0$, $w_0 \triangleq 0$),
    and \eqref{eq:proof:CRDF:SI@DEC:converse:CRDF} follows from \eqref{eq:CRDF:SI@DEC} for $x^t,w^t$ satisfying the 
    distortion and Markov constraints in \eqref{eq:CRDF:SI@DEC}.\footnote{If $w_t$ satisfies \eqref{eq: Markov for proof} it also satisfies the Markov constraint in \eqref{eq:CRDF:SI@DEC}.}
    %
    %
    %
    %
\end{IEEEproof}

\begin{remark}[SI causality]
\label{rem:KostinaHassibiWynerZiv_Informaion}
    Kostina and Hassibi \cite[Def.~3]{KostinaHassibi:SideInformationTracking} defined the (information) CRDF with decoder SI as 
    \begin{subequations}
    \label{eq:CRDF:SI@DEC:KostinaHassibi}
    \noeqref{eq:CRDF:SI@DEC:KostinaHassibi:cond}
    \begin{align}
             &R^{\mathrm{KH}}_{c}(D) \triangleq \limsup_{T \to \infty} 
         \inf_{\substack{ \causalKer{w^T}{x^T}, \{\hx_t(w^t,y^t)\} : 
         \\ \frac{1}{T} \sum_{t}\E{ \norm{x_t - \hx_t}^2 } \leq D 
         \\ \left(y^t,x^{t-1}\right)\markov\left(x^t,w^{t-1}\right)\markov w_t}}
            \!\frac{1}{T} \CDI{x^T\!}{w^T}{y^T} =  
    \label{eq:CRDF:SI@DEC:KostinaHassibi:cond}
         \\ &\limsup_{T \to \infty} 
         \inf_{\substack{ \causalKer{w^T}{x^T}, \{\hx_t(w^t,y^t)\} : 
         \\ \frac{1}{T} \sum_{t}\E{ \norm{x_t - \hx_t}^2 } \leq D 
         \\ \left(y^t,x^{t-1}\right)\markov\left(x^t,w^{t-1}\right)\markov w_t}}
            \!\frac{1}{T} \left\{ \DI{x^T\!}{w^T} - \DI{y^T\!}{w^T} \right\} .
    \label{eq:CRDF:SI@DEC:KostinaHassibi:diff}
    \end{align}
    \end{subequations}
    and prove that $R^{\mathrm{KH}}_{c}(D) = R^{\mathrm{both}}_{c}(D)$ in the Gaussian case \cite[Thm.~8]{KostinaHassibi:SideInformationTracking}.
    
    This definition can be viewed as an adaptation of the batch RDF with decoder \textit{non-causal} SI,  $R^\mathrm{NC}(D)$. Indeed, as $R^\mathrm{NC}(D) = R^\mathrm{both}(D)$ in the Gaussian (batch) case, 
    no improvement beyond the na\"ive bound of \lemref{lem:decSI:twoSided - information} is offered by  \eqref{eq:CRDF:SI@DEC:KostinaHassibi} for bounding the CRDF with decoder SI.
    
    Instead, we argue that better bounds result by relying on the technique of Weissman and El Gamal for batch RDF with decoder \textit{causal} SI, $R^\mathrm{C}(D)$.
    By comparing \eqref{eq:CRDF:SI@DEC} with \eqref{eq:CRDF:SI@DEC:KostinaHassibi:diff} the difference between the two bounds is $\frac{1}{T}\DI{y^T}{w^T} \geq 0$; as we shall claim in the sequel in \lemref{lem:decSI vs Two-Sided}, $\DI{y^T}{w^T} > 0$ in the Gaussian case, meaning that the bound offered by \thmref{thm:decSI:operational-information} is strictly better than that of \cite{KostinaHassibi:SideInformationTracking,Starvou:GaussRDF_SI}.
\end{remark}

\begin{remark}
    \label{rem:RateLoss_KostinaHassibiTwoSided}
    We note that without the Markov chain constraint in \eqref{eq:CRDF:SI@DEC:KostinaHassibi:cond} we could choose $w^t$ to be the $\hx^t$ that minimize \eqref{eq: Classical Causal RDF with SI:def}. Thus, in general, the inequality $R^{\mathrm{KH}}_{c}(D) \geq R^{both}_{c}(D)$ holds.
\end{remark}

\begin{lem}
\label{lem:decSI vs Two-Sided}
$R_c^{dec}(D)>R_c^{\mathrm{KH}}(D)$ whenever $R_c^{dec}(D) > 0$, and $R_c^{dec}(D) = R_c^{\mathrm{KH}}(D) = 0$ whenever $R_c^{dec}(D) = 0$.
\end{lem}

\begin{IEEEproof}[Proof sketch]
    The statement for $R_{c}^{dec}(D) = 0$ trivially follows from the non-negativity of the MI (see also \remref{rem:KostinaHassibiWynerZiv_Informaion}). 
    Assume 
    $R_c^{dec}(D) > 0$. Denote by $w_*^T$ the 
    $w^T$ that achieves the infimum in \eqref{eq:CRDF:SI@DEC}.
    Consider the following two cases.
    
    \textit{Case 1.}
        $w_*^T$ is jointly Gaussian with $x^T$ (and $y^T$) under 
        the limit superior in
        \eqref{eq:CRDF:SI@DEC}. 
        Then, $\underset{T \to \infty}{\limsup}\frac{1}{T}\DI{y^T}{w^T} > 0$ in \eqref{eq:CRDF:SI@DEC:KostinaHassibi:diff} \cite{StreamingWithFB:CNS, KostinaHassibi:SideInformationTracking}, and hence $R_c^{dec}(D) > R_c^{\mathrm{KH}}(D)$.
        
    \textit{Case 2.}
        $w_*^T$ is not jointly Gaussian with $x^T$ and $y^T$ under the limit superior 
        in \eqref{eq:CRDF:SI@DEC}.
        Denote by $w_G^T$ a jointly Gaussian vector with $x^T$ and $y^T$ that has the same joint second-order statistics with them as $w_*^T$.
        Then, we have 
        \begin{subequations}
        \label{eq:proof:WZ<EG}
        \begin{align}
            \DI{x^T}{w_*^T} &\geq \DI{x^T}{w_*^T} - \DI{y^T}{w_*^T}
            \label{eq:proof:WZ<EG:nonnegativity}
         \\ &= \CDI{x^T\!}{w_*^T}{y^T}
        \label{eq:proof:WZ<EG:CDI_definition}
         \\ &> \CDI{x^T\!}{w_G^T}{y^T} 
         \label{eq:proof:WZ<EG:UniquenessWZ}
        \end{align}
        \end{subequations}
        where \eqref{eq:proof:WZ<EG:nonnegativity} follows from the non-negativity of the DI, \eqref{eq:proof:WZ<EG:CDI_definition} is according to \eqref{eq:CRDF:SI@DEC:KostinaHassibi:diff} and \eqref{eq:proof:WZ<EG:UniquenessWZ} is from the uniqueness of the Gaussian solution of the problem \eqref{eq:CRDF:SI@DEC:KostinaHassibi} \cite{KostinaHassibi:SideInformationTracking}. Evaluating \eqref{eq:proof:WZ<EG} in $\underset{T \to \infty}{\limsup}$ yields the required result.
\end{IEEEproof}
\begin{corol}
\label{col:DecoderSI_TwoSidedSI_Compare}
    The following relations hold when~$R_{c}^{dec}(D) > 0$:
    \begin{align}
        \label{eq:InequalityChain_TwoSidedWEG}
        R_{c,op}^{dec}(D) \stackrel{(a)}\geq R_{c}^{dec}(D) \stackrel{(b)}> R_{c}^{\mathrm{KH}}(D) \stackrel{(c)}= R_{c}^{\mathrm{both}}(D) . 
    \end{align}
\end{corol}
\begin{IEEEproof}
    Steps $(a)$, $(b)$, and $(c)$ follow from 
    \thmref{thm:decSI:operational-information}, 
    \lemref{lem:decSI vs Two-Sided}, 
    and \cite[Thm.~8]{KostinaHassibi:SideInformationTracking} (see also \remref{rem:KostinaHassibiWynerZiv_Informaion}),
    respectively.
\end{IEEEproof}

\begin{corol}
The minimum distortion $D^\mathrm{dec}_{c,op}(D)$ of causal tracking of a Gauss--Markov source with causal SI over a memoryless channel with capacity $C$ 
is bounded from below~by
\begin{align}
    D_{c,op}^{dec}(C) 
    \geq \left(R_{c}^{dec}\right)^{-1}(C) 
    > \left(R_{c}^\mathrm{KH}\right)^{-1}(C) 
    = \left(R_{c}^\mathrm{both}\right)^{-1}(C) 
\nonumber
\end{align}

\end{corol}

\begin{IEEEproof}
    The proof is a simple adaptation of \cite[Thm.~2]{Causal-Noncausal_IEEEI2012}, \cite[Thm.~1]{MerhavShamai_WZ_GP_Separation}, which are in turn an adaptation of the necessity proof of the source--channel separation principle \cite[Thm~3.7]{ElGamalKimBook}; we outline it next.
    Denote the channel input and output at time $t$ by $a_t$ and $b_t$, respectively. Then, we have
    \begin{align}
         \label{eq:Prf:TrackingOverNoisyChannel}
         TR^{dec}_{c}(D) \stackrel{(a)}\leq \DI{x^T}{b^T} \stackrel{(b)}\leq \MI{x^T}{b^T} \stackrel{(c)}\leq TC
    \end{align}
%
%
    where $(a)$ is due to \defnref{def:Classical Causal RDF with SI @Rx} and noting that $b^T$ satisfies the conditions of $w^T$ in \eqref{eq:CRDF:SI@DEC}, $(b)$ holds since the DI is bounded from above by the MI, and $(c)$ is due to \cite[Eq. (31)]{Causal-Noncausal_IEEEI2012}.
    The proof then follows from
    \colref{col:DecoderSI_TwoSidedSI_Compare}, 
    by inverting the RDFs 
    and invoking their monotonicity \cite[Ch.~3]{ElGamalKimBook}.
\end{IEEEproof}
	
	\subsection{Adaptation for Control over Communication Channels}
\label{s: Discussion}


The CRDF of \defnref{def:Classical Causal RDF with SI @Rx} applies for a scenario where the SI is known only to the decoder. Consequently, the encoder cannot simulate the estimations $\hx^t$ of the decoder, and the auxiliary variables $w^t$ are not necessarily independent of the decoder outputs $\hx^{t-1}$. This, in turn, prevents using the CRDF of \defnref{def:Classical Causal RDF with SI @Rx} for control over communication channels (ConCom) as the estimation--control separation principle does not hold in this case \cite{Yuksel:AC2014:LQG:separation,TatikondaSahaiMitter}, \cite[Ch.~10]{YukselBasarBook} (see also references therein).

To circumvent this difficulty, we propose a genie-aided bound, in which, at time $t$, 
the SI signal $y_{t-1}$ is revealed to the encoder (via a genie), which can, therefore, construct $\hx_t$ and the innovation signal $x_t - \hx_t$. 
The 
genie-aided CRDF is 
\begin{align}
    \label{eq:CRDF_Genie}
\begin{aligned}
    &R^\mathrm{dec,g}_c(D) \triangleq 
    \\* & \limsup_{T \to \infty} \!\!\!\!
     \inf_{\substack{ \causalKer{w^T}{x^T,y^{T-1}}, \{\hx_t(w^t,y^t)\} : 
     \\ \frac{1}{T} \sum_{t}\E{ \norm{x_t - \hx_t}^2 } \leq D,
     \\ \left(y_t,x^{t-1}\right) \markov \left(x^t,w^{t-1},y^{t-1}\right) \markov w_t}}
     \!\!\!\! \frac{1}{T} \CDI{x^T}{w^T}{y^{T-1}} . \quad\:\:
\end{aligned}
\end{align}
Clearly, $R^{\mathrm{KH}}_{c}(D) \leq R^\mathrm{dec,g}_c(D) \leq R^\mathrm{dec}_c(D)$, where $R^\mathrm{dec,g}_c(D)$ 
%
may be utilized for the ConCom setup as the estimation--control separation principle extends to the genie-aided system. 
Furthermore, a simple adaptation of \lemref{lem:decSI vs Two-Sided} reveals that $R_c^{dec,g}(D)>R_c^{\mathrm{KH}}(D)$ whenever $R_c^{dec,g}(D) = 0$.

    \section{Numerical Simulations}
    \label{s: Numerical Evaluation - Gaussian Test Channel}
    
    We have seen in \lemref{lem:decSI vs Two-Sided} that $R_c^{dec}(D)$ gives a strictly tighter lower bound than that of $R_c^{both}(D)$ of \lemref{lem:decSI:twoSided - information} [and that of \eqref{eq:CRDF:SI@DEC:KostinaHassibi}] on the operational CRDF with decoder SI. 
    Unfortunately, carrying out the optimization in \eqref{eq:CRDF:SI@DEC} and finding an explicit solution is difficult and is yet to be determined even for the simpler memoryless batch, in which it reduces to the single-letter optimization problem in \eqref{eq:RDF:WeissmanElGamal}.
    
    Following \cite{WeissmanElGamal_FiniteLookAhead}, we consider a Gaussian test channel---$w_t = x_t + z_t$, where $z_t$ is a zero-mean AWGN of variance $\sigma_z^2$ in lieu of the infimum in \eqref{eq:CRDF:SI@DEC} and evaluate the expression for this choice. We shall further show that Gaussian test channels are suboptimal meaning that Case 2 prevails in the proof of \lemref{lem:decSI vs Two-Sided}.
    We denote the minimum mean square errors (MMSEs) given $w^t$ and given $(y^t, w^t)$ by
    \begin{align}
    \label{eq:distortion-definition:numerics}
        D_t &= \E{ \left(x_{\tind} - \hat{x}_{\tind}(y^{\tind},w^{\tind})\right)^2 }, 
        &\tD_t &= \E{ \left(x_{\tind} - \hat{x}_{\tind}(w^{\tind})\right)^2 } . 
    \end{align}

    
    First, note that $R_1$ equals the channel capacity of a power constrained AWGN channel \cite{CoverBook}:
    \begin{align}
        \label{eq: Rate, initial}
        R_1 = I(x_{1};w_{1}) = \frac{1}{2}\log\left(1 + \frac{\sigma^{2}_{v}}{\sigma^{2}_{z}}\right) ,
    \end{align}
    and $D_1 = \left. \sigma_v^2 \middle\| \sigma_n^2 \middle\| \sigma_z^2 \right.$. 
    By substituting it in \eqref{eq: Rate, initial}, 
    we arrive at
    \begin{align}
        \label{eq: RDF, initial}
            R_1 = \frac{1}{2}\log\left(\frac{\sigma^2_{v}}{D_{1}} - \frac{\sigma^2_{v}}{\sigma^2_{n}}\right) .
    \end{align}
    Since rate--distortion curves must be convex and non-negative \cite[Ch.~10]{CoverBook}, 
    we clip $R_1$ of \eqref{eq: RDF, initial} at 0 and take its lower convex envelope to be the rate--distortion curve $R_1(D_1)$. 

   \begin{figure}[t]
		\vspace{-.65\baselineskip}
		\begin{subfigure}[t]{\columnwidth}
    		\centering
		    \includegraphics[width=\columnwidth]{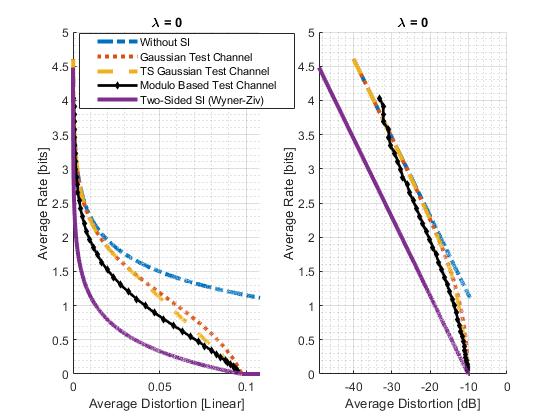}
		\vspace{-1.3\baselineskip}
		    \caption{$\lambda = 0$}
		\end{subfigure}
		\begin{subfigure}[t]{\columnwidth}
    		\centering
		    \includegraphics[width=\columnwidth]{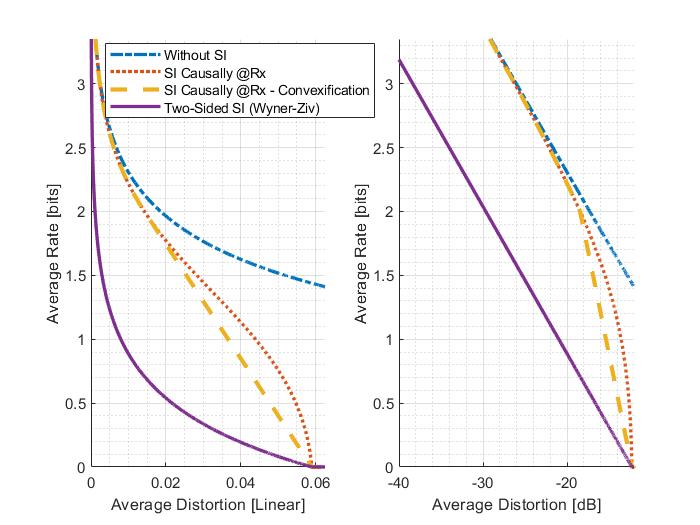}
		\vspace{-1.3\baselineskip}
		    \caption{$\lambda = 0.9$}
		\end{subfigure}
		\caption{Information average rate versus the average distortion for no SI, two-sided SI, and causal decoder SI with a Gaussian test channel $w_t = x_t + n_t$ with $\sigma_n = 1/3$ for $\lambda = 0, 0.9$.
		We use a uniform distortion allocation 
		$D_1 = \cdots = D_T = D$ in all the curves and $T = 2048$.}
		\label{fig:RDFs}
	\end{figure}
	
    By putting forth the the process dynamics \eqref{eq:Source Gauss Markov} and pedestrian MMSE estimation arguments 
     we arrive at
     \begin{align}
        \label{eq: distortion evolution}
        \!\! D_{t+1} &= \left. \sigma_n^2 \middle\| \sigma_z^2 \middle\| \left( \lambda^2 D_t + \sigma_v^2 \right) \right. \!,
        & \!\! \tD_{t+1} &= \left. \sigma_z^2 \middle\| \left( \lambda^2 \tD_t + \sigma_v^2 \right) \right.\! . \ \ \ 
    \end{align}
     
     By defining $R_t(\tD_t) \triangleq \half \log \left( \lambda^2 + \frac{\sigma_v^2}{\tD_t} \right)$ for $t > 1$, \eqref{eq: RDF, initial}, 
     and using \cite[Proof of Corol.~2]{StreamingWithFB:CNS},  \cite[Thm.~2]{KostinaHassibi:SideInformationTracking}, 
     we have ($\tD_0 = 0$)
     \begin{align}
    \label{eq: Overall Rate Definition}
    \begin{aligned}
        \DI{x^T}{w^T} 
        &= \frac{1}{2}\log\left(\frac{\sigma^{2}_{v}}{\tD_{1}}\right) + \frac{1}{2}\sum_{\tind=2}^{T}\log\left(\lambda^2 + \frac{\sigma^{2}_{v}}{\tD_{t}}\right) 
     \\ &\triangleq \sum_{\tind=1}^{T} R_{\tind}(\tD_t) .
    \end{aligned}
    \end{align}

    Using the definition of $R_t(\tD_t)$ and 
    %
    \eqref{eq: distortion evolution}, we obtain
    \begin{align}
        \label{eq: recursive d tilde}
         &\tD_{t+1} = \left. \sigma^2_z \middle\| \sigma^2_v \left( 1 - \lambda^2 2^{-2R_t} \right)^{-1} \right. .
    \end{align}    
    
    And by equating \eqref{eq: recursive d tilde} with $\tD_{t+1}$ of the definition of $R_{t+1}(\tD_{t+1})$ we attain 
    \begin{align}
        \label{eq: recursive sigma_z}
        \sigma^2_z = \frac{\sigma^{2}_{v}}{2^{2R_{\tind+1}} - 1 - \lambda^{2}\left(1 - 2^{-2R_{\tind}}\right)} \:.
    \end{align}
     Substituting \eqref{eq: recursive sigma_z} into the recurssion of $D_{t+1}$ \eqref{eq: distortion evolution} we arrive at the 
     recursive description: 
     \begin{align}
    \label{eq: recursive rate distortion}
        R_{t+1}=\frac{1}{2}\log\left(\frac{\sigma^{2}_{v}}{D_{t+1}}-\frac{\sigma^{2}_{v}}{\sigma^{2}_{n}}- \frac{\sigma^{2}_{v}}{\lambda^{2}D_{t}+\sigma^{2}_{v}} +\lambda^{2}\left(1-2^{-2R_{t}}\right)+1\right) \!.
    \end{align}   
    
    By substituting \eqref{eq: recursive rate distortion} and \eqref{eq: RDF, initial} into \eqref{eq: Overall Rate Definition} we get an expression for the average rate. 
    
    The steady-state solution for \eqref{eq: recursive rate distortion} is given by 
    \begin{align}
        R = \half \log^+ \left( \lambda^2 + \frac{\sigma_v^2}{\tD} \right) ,
    \end{align}
    where $\tD$ is the positive solution of the quadratic equation

    \begin{align}
        \lambda^2 \tD^2 + \left[ \sigma_v^2 + \left( 1 - \lambda^2  \right) \sigma_z^2 \right] \tD - \sigma_v^2 \sigma_z^2 = 0 , 
    \end{align}
    whereas the distortion is given by the positive solution of the quadratic equation

    \begin{align}
        \lambda^2 D^2 + \left[ \sigma_v^2 + \left( 1 - \lambda^2  \right) \left( \sigma_z^2 \middle\| \sigma_n^2 \right) \right] D - \sigma_v^2 \left( \sigma_z^2 \middle\| \sigma_n^2 \right) = 0 . 
    \end{align}
    
    This curve is not convex meaning that the optimal test channel in \eqref{eq:CRDF:SI@DEC} is not Gaussian. Consequently, by convexifying (corresponding to time-sharing with $R=0$), we improve this curve.\\
    We further consider one-dimensional modulo-based mappings 
    \cite{JointWZ-WDP,Tuncel_ZeroDelayJSCCwithWZ} and show that they outperform the TS curve of \cite{WeissmanElGamal_FiniteLookAhead}. 
    
    Following \cite{lev2020schemes}, we further consider one-dimensional modulo-based mappings 
    \cite{Tuncel_ZeroDelayJSCCwithWZ,JointWZ-WDP}, whose equivalent test channel is given by 
    \begin{align}
        \label{eq:TestChannel_Modulo}
        w_t = \alpha [x_t]_\Delta + \beta \left(x_t - [x_t]_\Delta\right) + z_t
    \end{align}
    with $[x_t]_\Delta \triangleq x_t - \Delta \cdot \mathrm{round}\left(x_t/\Delta\right)$ and $\alpha,\beta,\Delta$ are chosen such that the power of $x_t$ is preserved.

    

    
    We plot the resulting (convexified and 1D modulo) curves

We plot the achievable curves for $\lambda = 0,0.9$, $\sigma_{n} = 1/3, \sigma_v = 1$  and compare them to 
$R^\mathrm{KH}_c(D) \equiv R^\mathrm{both}_c(D)$ of \eqref{eq: Classical Causal RDF with SI:Gauss}, \eqref{eq:CRDF:SI@DEC:KostinaHassibi} and $R_c(D)$ of \eqref{eq: Classical Causal RDF:Gauss};
clearly, the TS Gaussian test-channel \cite{WeissmanElGamal_FiniteLookAhead} and modulo-based \cite{Tuncel_ZeroDelayJSCCwithWZ} curves lie between the latter two.
 We further notice that the modulo-based mappings \cite{Tuncel_ZeroDelayJSCCwithWZ} achieve better performance than that of the TS Gaussian test-channel solution of \cite{WeissmanElGamal_FiniteLookAhead}.

    \bibliographystyle{IEEEtran}
    \bibliography{myBib}
\end{document}